\documentstyle[12pt,a4,epsfig]{article}
\begin{document}
\begin{flushright}
BI-TP 96/11 \\
hep-ph/9603283 
\end{flushright}
\vspace{1cm}
\begin{center}
\Large{\bf
Discovering Non-Abelian Weak Couplings and an\\
Anomalous Magnetic Dipole Moment
of the \boldmath $W^{\pm}$\unboldmath\\
at LEP~2}
\end{center}
\vspace{1cm}
\begin{center}
{\bf I. Kuss\footnote{\normalsize e-mail: kuss@hrz.uni-bielefeld.de},
D. Schildknecht}\\
Universit\"at Bielefeld\\
Fakult\"at f\"ur Physik\\
Postfach 10 01 31\\
D--33501 Bielefeld, Germany\\
\vspace{.5cm}
Revised Version, May 1996
\end{center}
\vspace{.5cm}
\renewcommand{\thefootnote}{\fnsymbol{footnote}}
\setcounter{footnote}{2}
\begin{abstract}
\noindent 
In view of the forthcoming results on $W^{\pm}$ pair production at LEP~2,
 we emphasize that {\em direct empirical evidence} on non-trivial
properties of the weak vector bosons can be obtained with relatively
limited integrated luminosity.
An integrated luminosity of $10\;\mathrm{pb}^{-1}$ at
$\sqrt{s}=175$ GeV will be sufficient to provide direct experimental
evidence for non-vanishing self-couplings of non-Abelian type
among the weak vector bosons.
An integrated luminosity of $100\;\mathrm{pb}^{-1}$ at
$\sqrt{s}=175$ GeV will provide direct evidence for
the existence of an anomalous magnetic dipole moment
of the charged vector bosons
$W^{\pm}$.
\end{abstract}
\section*{}

With respect to properties of the electroweak vector bosons, there 
are two salient predictions of the standard electroweak theory
\cite{sm}
which lack direct experimental confirmation so far: the non-Abelian
coupling between the members of the $SU(2)$ triplet of vector bosons,
and, closely connected to it, the anomalous magnetic 
dipole moment coupling of
the charged vector bosons, $W^{\pm}$. From the detailed analysis of the
LEP~1 precision data there is strong {\em indirect} evidence
\cite{delta_y} that indeed the standard couplings are realized in nature.
The production of $W^+W^-$ at LEP~2 will nevertheless open up a
new domain of investigations by providing {\em direct} tests of the
non-Abelian structure and the anomalous magnetic dipole moment prediction.
\par
Limited luminosity will restrict the possibility of precision
measurements of the trilinear $\gamma W^+W^-$ and the $Z^0 W^+W^-$
couplings at LEP~2. It is the purpose of the present note to point out
that a rather small integrated luminosity at LEP~2, however, will
nevertheless be sufficient to provide empirical evidence for the
existence of a non-Abelian trilinear 
coupling among the massive vector bosons and
for the existence of an anomalous magnetic dipole moment of the $W^{\pm}$.\par
\vspace{0.3cm}
We start from the phenomenological Lagrangian for trilinear vector boson
couplings widely used in the
simulation of future data \cite{BKRS}\footnote{We restrict ourselves to
dimension-four, P- and C-conserving interactions.},
\begin{eqnarray}
{\cal L}_{int}&=&-ie[A_\mu(W^{-\mu\nu}W^+_\nu-W^{+\mu\nu}W^-_\nu)
+F_{\mu\nu}W^{+\mu}W^{-\nu}]\nonumber\\
&&-iex_\gamma F_{\mu\nu}W^{+\mu}W^{-\nu}\nonumber\\
&&-ie(\frac{c_W}{s_W}+\delta_Z)[Z_\mu(W^{-\mu\nu}W^+_\nu-W^{+\mu\nu}W^-_\nu)
+Z_{\mu\nu}W^{+\mu}W^{-\nu}]\nonumber\\
&&-iex_Z Z_{\mu\nu}W^{+\mu}W^{-\nu}.
\label{xdellag}
\end{eqnarray} 
It contains the three arbitrary parameters $x_\gamma$, $\delta_Z$ and
$x_Z$ and reduces to the standard form for $x_\gamma=x_Z=\delta_Z=0$.
The parameter $x_\gamma$ is directly related to the anomalous magnetic dipole
moment $\kappa_\gamma$ of the $W^{\pm}$ via
\begin{equation}
x_\gamma\equiv \kappa_\gamma-1,
\label{kappagam}
\end{equation}
where $\kappa_\gamma=1$ is the value in the standard model.
This value follows from the linear realization of the $SU(2)\times U(1)$
symmetry in conjunction with the restriction to dimension-4 terms as
embodied in the standard electroweak theory \cite{sm}. Realizing $SU(2)\times
U(1)$ symmetry non-linearly, or else removing the restriction to dimension-4
terms in the linearized form, removes the restriction $\kappa_{\gamma}=1$,
e.g. \cite{gks2}. We also note that $\kappa_{\gamma}=1$ corresponds to a
gyromagnetic ratio, $g$, of the $W^{\pm}$ of magnitude $g=2$, in units
of the particle's Bohr-magneton $e/(2M_W)$, 
while $\kappa_{\gamma}=0$ corresponds to $g=1$ as obtained
for a classical rotating charge distribution. In general $g=1+\kappa_{\gamma}$.
The value of $g=2$ puts
the electromagnetic properties of the $W^{\pm}$ in close analogy to the
$g=2$ value of the spin-$1/2$ Dirac theory.
\vspace{0.3cm}
\par In order
to relate $\delta_Z$ and $x_Z$ to the parameters in the Lagrangian before
diagonalization of the neutral sector, it is advantageous to rewrite 
(\ref{xdellag}) in terms of unmixed neutral vector boson fields. For
the present purpose it is useful to describe mixing in terms of 
current-mixing \cite{hs}, i.e a mixing term between the third component of the
weak isotriplet, $\tilde{W}_3^{\mu\nu}$, and the photon field tensor,
which is denoted by 
$\tilde{F}^{\mu\nu}\equiv\partial^{\mu}\tilde{A}^{\nu}-\partial^{\nu}
\tilde{A}^{\mu}$ in order to 
discriminate $\tilde{F}^{\mu\nu}$ from the physical photon field tensor
$F^{\mu\nu}$ emerging upon
diagonalization. For details we refer to \cite{kmss}. Carrying out the
transformtion to the unmixed fields,
\begin{equation}
\pmatrix{A^{\mu}\cr Z^{\mu}\cr}=\pmatrix{1&s_W\cr 0&c_W\cr}\pmatrix{\tilde{A}^{\mu}\cr
\tilde{W}_3^{\mu}\cr},
\label{trafo1}\end{equation}
the Lagrangian (\ref{xdellag}) becomes
\pagebreak
\begin{eqnarray}
{\cal L}_{int}&=&\frac{\hat{g}}{2}\vec{W}_{\mu\nu}\cdot
(\vec{W}^\mu\times\vec{W}^\nu)\cr
&&-i\hat{f}\tilde{W}^3_{\mu\nu}W^{+\mu}W^{-\nu}\cr
&&-ie\left[\tilde{A}_\mu(W^{-\mu\nu}W^+_\nu-W^{+\mu\nu}W^-_\nu)
+\kappa_\gamma \tilde{F}_{\mu\nu}W^{+\mu}W^{-\nu}\right],\label{Lint}
\end{eqnarray}
where the third component of $\vec{W}$ is understood to be $\tilde{W}_3$.
The relations between $\hat{g}$, $\hat{f}$, $\kappa_\gamma$ and
$x_\gamma$, $x_Z$ and $\delta_Z$ are given by (\ref{kappagam}) and
\begin{eqnarray}
\delta_Z&=&\frac{\hat{g}}{ec_W}-\frac{1}{s_Wc_W},\cr
x_Z&=&\frac{\hat{f}}{ec_W}-(\kappa_\gamma-1)\frac{s_W}{c_W}.
\label{rela1}\end{eqnarray}
The standard model thus corresponds to $\hat{g}=\frac{e}{s_W}$,
$\hat{f}=0$ and $\kappa_\gamma=1$.\par
Let us consider the Lagrangian (\ref{Lint}) in some detail. It contains
\cite{kmss}:
\begin{itemize}
\item An $SU(2)$ symmetric interaction term of non-Abelian form with
arbitrary strength $\hat{g}$ which coincides with the standard term for
the special choice $\hat{g}=\frac{e}{s_W}$. Note that this term is
contained in the kinetic term for the $\vec{W}$ fields,
\begin{equation}
-\frac{1}{4}\vec{W}^{\mu\nu}\vec{W}_{\mu\nu}
=-\frac{1}{2}W^{-\mu\nu}W^+_{\mu\nu}-\frac{1}{4}\tilde{W}_3^{\mu\nu}
\tilde{W}^3_{\mu\nu},
\label{wkin}\end{equation}
provided we use the non-Abelian field tensor,
\begin{equation}
\vec{W}^{\mu\nu}=\partial^{\mu}\vec{W}^{\nu}-\partial^{\nu}\vec{W}^{\mu}
-\hat{g}\vec{W}^{\mu}\times\vec{W}^{\nu}.
\label{nonab}
\end{equation}
For $\hat{g}=0$, the kinetic term reduces to a triplet of Abelian vector
boson fields\footnote{The local Abelian symmetry, $W_{\mu}^i\to
W_{\mu}^i+\partial_{\mu}\alpha^i(x)$,  is obviously 
broken by the mass terms
for the vector bosons.}. Accordingly, 
the case of $\hat{g}=0$ may be referred to as
the Abelian triplet model.
\item A term of strength $\hat{f}$ which violates $SU(2)$ symmetry
independently of the presence of the photon field. Imposing the
constraint of no intrinsic $SU(2)$ violation\cite{kmss}, i.e. $SU(2)$ symmetry,
when electromagnetism in (\ref{Lint}) is absent, we have to require
$\hat{f}=0$, i.e. 
\begin{equation}
x_Z=-\frac{s_W}{c_W}x_\gamma.
\label{su2c_rel}
\end{equation}
This requirement is
abstracted from its empirical validity in the vector boson mass term and
is sometimes called custodial $SU(2)_c$ symmetry.
\item The two terms describing the electromagnetic
interactions. The term containing $\tilde{A}^{\mu}$ in (\ref{Lint}) 
is simply obtained
by applying the minimal substitution principle, $\partial^{\mu}\to
\partial^{\mu}+ie\tilde{A}^{\mu}Q$, $Q$ being the charge operator,
$QW^\pm=\pm W^\pm$ and $QW_3=0$, to the derivatives in the kinetic term 
(\ref{wkin}) of the $\vec{W}$
triplet. 
The anomalous magnetic moment term $\kappa_\gamma$, $\kappa_\gamma\neq 0$
in (\ref{Lint}), on the other hand,
need not necessarily be present, even though it is not excluded by the
principle of minimal substitution \cite{lees}. In fact, adding the
$SU(2)$-invariant total derivative term
\begin{eqnarray}
{\cal L}_{\mathrm{tot}}&=&-\frac{\kappa_{\gamma}}{2}\partial_{\mu}
\left(W^{+\nu}\partial_{\nu}W^{-\mu}-W^{+\mu}\partial_{\nu}W^{-\nu}\right.\cr
&&\;\;\;\;\;\;\;\;\;\left.+\;W^{-\nu}\partial_{\nu}W^{+\mu}-W^{-\mu}\partial_{\nu}W^{+\nu}\right.\cr
&&\;\;\;\;\;\;\;\;\;\left.+\;\tilde{W}_3^{\nu}\partial_{\nu}\tilde{W}_3^{\mu}-\tilde{W}_3^{\mu}\partial_{\nu}\tilde{W}_3^{\nu}\right)
\label{Ltot}\end{eqnarray}
to the kinetic term (\ref{wkin}) and carrying out the minimal
substitution prescription implies the electromagnetic interaction (\ref{Lint}) with
an arbitrary value of $\kappa_{\gamma}$.
\par
We note that both the charge
coupling, $e$, as well as the magnetic moment coupling, $\kappa_\gamma$,
appearing in connection with the unmixed field $\tilde{A}^{\mu}$,
$\tilde{F}^{\mu\nu}$ in (\ref{Lint}), 
are identical to the couplings in the Lagrangian
(\ref{xdellag}) for the physical fields, $A^{\mu}$, $F^{\mu\nu}$.
This is in contrast to the case of the massive neutral vector boson,
where the third component of the triplet in (\ref{Lint}) couples with a
strength $\hat{g}$ different from the strength of the $Z$ coupling,
$e\left(\frac{c_W}{s_W}+\delta_Z\right)$, in (\ref{xdellag}).
\end{itemize}
\par
We note in passing that the mentioned $SU(2)$-symmetry properties
of the terms multiplied by $\hat{g}$ and $\hat{f}$ in (\ref{Lint})
are not a unique feature of the
$\tilde{A}\tilde{W}_3$ basis.
The physical interpretation of the parameters $\hat{g}$
and $\hat{f}$ is thus not only valid in this basis.
Indeed, expressing (\ref{xdellag}) in terms of the more conventional
$BW_3$ basis via
\begin{equation}
\pmatrix{A^\mu\cr Z^\mu\cr}=\pmatrix{c_W&s_W\cr-s_W&c_W\cr}\pmatrix{B^\mu\cr
W_3^\mu\cr},\label{trafo2}\end{equation}
one obtains the Lagrangian in the form
\begin{eqnarray}
{\cal L}_{int}&=&\frac{\hat{g}}{2}\vec{W}_{\mu\nu}\cdot
(\vec{W}^\mu\times\vec{W}^\nu)\cr
&&-i\hat{f}W^3_{\mu\nu}W^{+\mu}W^{-\nu}\cr
&&-i\left[\left(\frac{e}{c_W}-\hat{g}\frac{s_W}{c_W}\right)
B_{\mu}(W^{-\mu\nu}W^+_\nu-W^{+\mu\nu}W^-_\nu)\right.\cr
&&\;\;\;\left.+\left(\frac{e}{c_W}\kappa_\gamma-(\hat{g}+\hat{f})
\frac{s_W}{c_W}\right) B_{\mu\nu}W^{+\mu}W^{-\nu}\right].\label{LBW3}
\end{eqnarray}
This form of the Lagrangian coincides with (\ref{Lint}) in the first two
terms apart from $\tilde{W}_3^{\mu}$ being replaced by $W_3^{\mu}$.
This is a consequence of the fact that the second columns of the matrices
in (\ref{trafo1}) and (\ref{trafo2}) are identical.
Obviously, however, the $B$-dependent terms in (\ref{LBW3})
differ from the $\tilde{A}$ terms in (\ref{Lint}).
\par
In summary, it is a characteristic feature of a non-Abelian theory to
contain a coupling $\hat{g}\neq 0$. Experimental evidence for
$\hat{g}\neq 0$ thus provides evidence for a non-Abelian structure of
the interactions among the members of the $\vec{W}$ triplet and rules
out a theory based on three Abelian vector boson fields.\par
It is a second characteristic feature of the non-Abelian nature
of the standard model interactions to contain
an anomalous magnetic dipole moment term of definite strength, 
$\kappa_\gamma=1$. While a
precision measurement of $\kappa_\gamma$ will be difficult, empirical
evidence for $\kappa_\gamma > 0$ will nevertheless provide first direct
experimental evidence for the existence of an
anomalous magnetic dipole moment of the $W^{\pm}$ vector bosons.\par
We turn to the experimental search for a non-Abelian coupling,
$\hat{g}\neq 0$,
and the search for an anomalous magnetic dipole moment, $\kappa_\gamma\neq 0$.

We note that the present most stringent direct bound to the 
$WW\gamma$-coupling $x_\gamma$ 
\cite{Tevatron} is $-1.6 < x_\gamma < 1.8$ 
at 95\% CL\footnote{These bounds depend on the assumption of a form factor
in the $WW\gamma$ coupling. 
The bounds
to the $WWZ$ couplings reported in \cite{Tevatron} are based on specific
model assumptions which are inconsistent with the present work.},
corresponding to $-0.6 < \kappa_\gamma < 2.8$.

We consider a measurement of the total cross section of 
$e^+ e^-\to W^+ W^-$ at LEP~2 with a cut $|\cos\theta|<0.98$
on the scattering angle.
Formulae for the cross section in terms of the parameters
$x_\gamma,x_Z$ and $\delta_Z$ have been given in \cite{BKRS}.
We assume that the decay mode 
\begin{equation}
e^+ e^-\to W^+ W^-\to l^\pm \nu_l\;+\;\mbox{2 jets},
\label{epemtojets}
\end{equation}
will be detected at LEP~2,
where the lepton $l^\pm$ can be either an electron (positron) 
or a muon (anti-muon).
The branching ratio for this decay mode is 29.6\%.
We assume that future data are identical to the standard model
predictions and
calculate the lines of constant 
$\chi^2=4$ (86.5\% CL) and $\chi^2=1$ (39.4\% CL)
in the $x_\gamma$-$\delta_Z$-plane according to\footnote{
We assume that the measurement is the outcome of
a random sample of a
Gaussian distribution with mean value $N(x_{\gamma},\delta_Z)$ such that
the statistical error is given by $\sigma=\sqrt{N(x_{\gamma},\delta_Z)}$.
If one uses the error estimated from the experiment, $\sigma=\sqrt{N_{SM}}$,
instead of the statistical error in the denominator of (\ref{chi2}),
the values of the parameter pair ($x_{\gamma},\delta_Z$) which have
$\chi^2=4$ or $\chi^2=1$ change only little provided the number of measured
events, $N_{SM}$, is much greater than one, $N_{SM}\gg 1$.}
\begin{equation}
\chi^2\equiv
\frac{\left(N_{SM}-N(x_{\gamma},\delta_Z)\right)^2}
{N(x_{\gamma},\delta_Z)}.
\label{chi2}
\end{equation}
In (\ref{chi2}), $N_{SM}$ and $N(\delta_Z,x_\gamma)$ denote the number of events
in the SM and in the alternative model with the $SU(2)$ constraint
(\ref{su2c_rel}) and $\delta_Z,x_{\gamma}\neq 0$, respectively.
\par
\begin{figure}
\begin{center}
\psfig{file=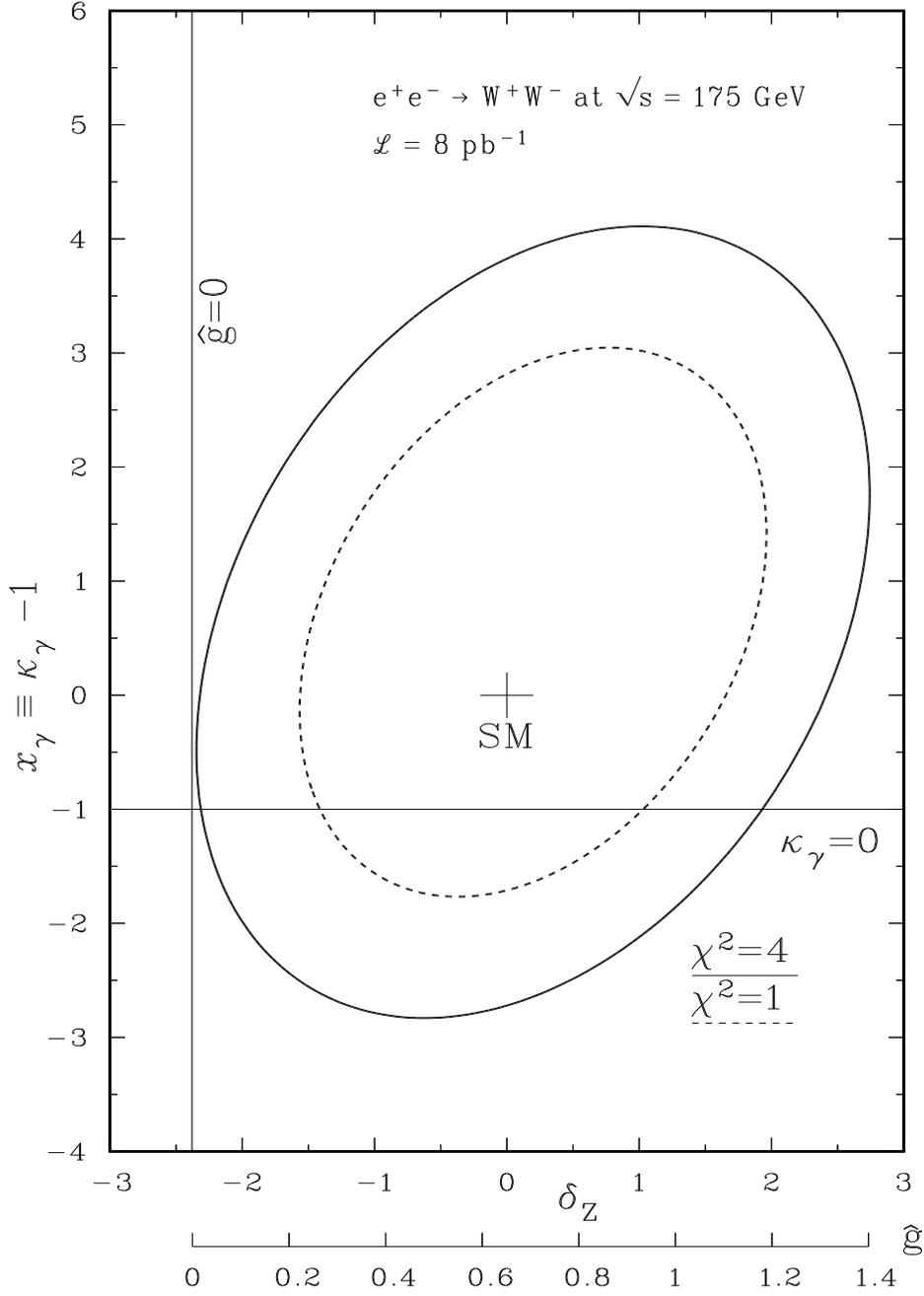,width=15cm,height=19cm}
\end{center}
\caption{Detecting a non-Abelian vector-boson coupling, 
\protect{$\hat{g}\neq 0$}, at LEP~2.}
\label{fig1}
\end{figure}

In Figure \ref{fig1} we show the $1\sigma$ and $2\sigma$ contours in the
$(\delta_Z,x_\gamma)$ plane for an integrated luminosity of 
${\cal L}=8\;\mathrm{pb}^{-1}$, 
at $\sqrt{s}=$ 175 GeV. We see that this very small value of ${\cal L}$,
corresponding to a few weeks of running at LEP~2, is sufficient to detect a
genuine non-zero non-Abelian coupling, $\hat{g}\neq 0$, at the $2\sigma$
level. Likewise, a vanishing anomalous magnetic moment, $\kappa_{\gamma}=0$,
in conjunction
with an Abelian 
$\vec{W}$ triplet, $\hat{g}=0$, is excluded. We note
that similar results can be obtained 
with ${\cal L}=20\;
\mathrm{pb}^{-1}$ at $\sqrt{s}=170$ GeV and with ${\cal L}=2.2\;
\mathrm{pb}^{-1}$ at $\sqrt{s}=190$ GeV.\par
\begin{figure}
\psfig{file=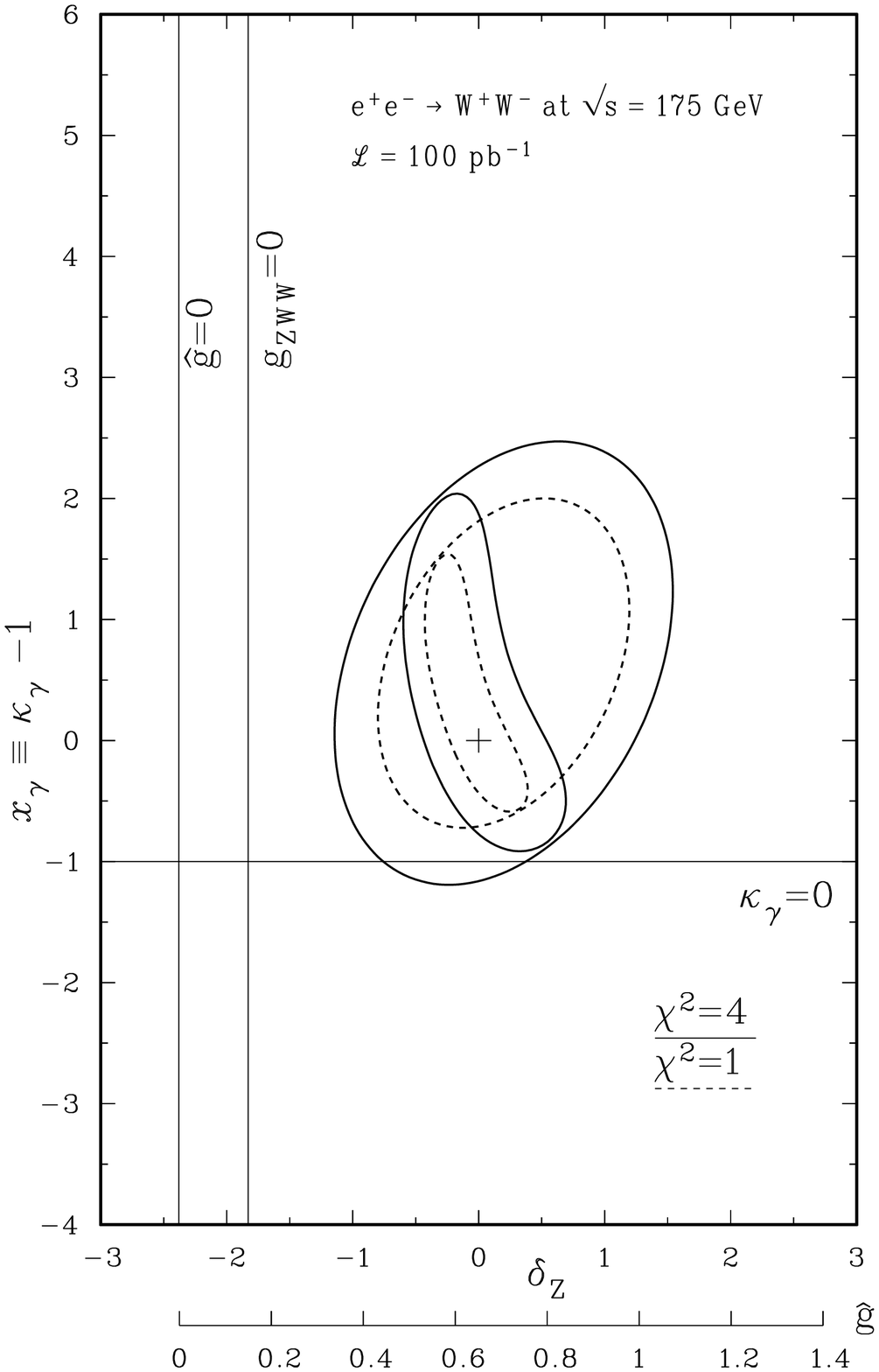,width=15cm,height=19cm}
\caption{Detecting a non-zero anomalous magnetic dipole moment,
\protect{$\kappa_\gamma\neq 0$}, of the
\protect{$W^{\pm}$} at LEP~2.} 
\label{fig2}
\end{figure}

To detect a non-vanishing magnetic dipole moment, $\kappa_\gamma\neq 0$, in the
presence of a non-vanishing non-Abelian coupling, $\hat{g}\neq 0$, needs
a somewhat higher integrated luminosity. The result in Figure \ref{fig2}
corresponds to a luminosity
of ${\cal L} = 100\;\mathrm{pb}^{-1}$ at $\sqrt{s}=175$ GeV, thus
providing direct evidence for a non-vanishing anomalous magnetic dipole moment of the $W^\pm$,
$\kappa_\gamma > 0$. Since a sufficient number of standard events,
$N_{SM}=519$, is expected in this case, the events can be arranged in 6 bins
equidistant over the scattering angle $\cos\theta$, leading to the
result also presented in Figure \ref{fig2}. In the same Figure we also
show the theoretical prediction corresponding to a vanishing $ZW^+W^-$
coupling, $g_{ZWW}$, corresponding to $\delta_Z=-\frac{c_W}{s_W}$ in 
(\ref{xdellag}), and $\hat{g}=es_W$, which is similarly ruled out.\par
In conclusion, after a few weeks of running at $\sqrt{s}=175$ GeV at LEP~2,
definite direct evidence for the existence of a genuine, non-vanishing
coupling among the members of the $W^\pm,W^0$ triplet, characteristic of
a non-Abelian structure, can be
obtained. Likewise, after 7 months of running at LEP~2, definite
evidence for a non-vanishing anomalous magnetic dipole moment of the charged
vector bosons may be expected.

\section*{Acknowledgement}
The authors thank S. Dittmaier for useful discussions.


\unitlength1cm
\newpage

\newpage

\begin{thebibliography}{99}
\bibitem{sm} S. L. Glashow, Nucl. Phys. {\bf 22} (1961) 579;
S. Weinberg, Phys. Rev. Lett. {\bf 19} (1967) 1264; A.
Salam, Proc. 8th Nobel Symposium, ed. N. Svartholm (Almquits and
Wiksells, Stockholm, 1968), p. 367
\bibitem{delta_y} S. Dittmaier, D. Schildknecht and G. Weiglein,
BI-TP 95/31, hep-ph/9510386, to be published in Nucl. Phys {\bf B}
\bibitem{BKRS}M. Bilenky, J. L. Kneur, F. M. Renard and D. Schildknecht,
Nucl. Phys. {\bf B409} (1993) 22
\bibitem{gks2} C. Grosse-Knetter, I. Kuss and D. Schildknecht, Phys.
Lett. {\bf B358} (1995) 87
\bibitem{hs} P. Q. Hung and J. J. Sakurai, Nucl. Phys. {\bf B143} (1978) 81
\bibitem{kmss}J. Maalampi, D. Schildknecht and K.H. Schwarzer,
Phys. Lett. {\bf B166} (1986) 361;
M. Kuroda, J. Maalampi, D. Schildknecht and K. H. Schwarzer,
Nucl. Phys. {\bf B284} (1987) 271; Phys. Lett. {\bf B190} (1987) 217
\bibitem{lees} H. C. Corben and J. Schwinger, Phys. Rev. {\bf 58}
(1940) 953; T. D. Lee, Phys. Rev. {\bf 140} (1965) B967
\bibitem{Tevatron} The CDF Collaboration, Phys. Rev. Lett {\bf 75} (1995)
1017; The D0 Collaboration, Phys. Rev. Lett {\bf 75} (1995) 1023 and 1034
\end{thebibliography}
\end{document}